\documentclass[twocolumn,amsmath,amssymb,superscriptaddress]{revtex4-2}

\usepackage{xcolor}

\usepackage{graphicx}

\usepackage{mathrsfs}

\newcommand{\ud}{\mathrm{d}}

\graphicspath{ {./figures/} {./}}

\begin{document}

\title{Modulational instability in randomly dispersion-managed fiber links}

\author{Andrea Armaroli} 

\affiliation{Univ.~Lille,  CNRS,  UMR  8523-PhLAM-Physique  des  Lasers  Atomes  et  Mol\'ecules,  F-59000  Lille,  France}

\author{Guillaume Dujardin} 

\affiliation{Univ.~Lille, Inria, CNRS, UMR 8524 - Laboratoire Paul Painlev{\'e}, F-59000 Lille, France}

\author{Alexandre Kudlinski} 
\affiliation{Univ.~Lille,  CNRS,  UMR  8523-PhLAM-Physique  des  Lasers  Atomes  et  Mol\'ecules,  F-59000  Lille,  France}

\author{Arnaud Mussot} 

\affiliation{Univ.~Lille,  CNRS,  UMR  8523-PhLAM-Physique  des  Lasers  Atomes  et  Mol\'ecules,  F-59000  Lille,  France}

\author{Stephan De Bi{\`e}vre} 

\affiliation{Univ.~Lille, CNRS, Inria, UMR 8524 - Laboratoire Paul Painlev{\'e}, F-59000 Lille, France}

\author{Matteo Conforti} 

\affiliation{Univ.~Lille,  CNRS,  UMR  8523-PhLAM-Physique  des  Lasers  Atomes  et  Mol\'ecules,  F-59000  Lille,  France}

\begin{abstract}
We study modulational instability in a dispersion-managed system where the sign of the group-velocity dispersion is changed at uniformly distributed random distances around a reference length. An analytical technique is presented to estimate the instability gain from the linearized nonlinear Schr{\"o}dinger equation, which is also solved numerically. The comparison of numerical and analytical results confirms the validity of our approach. Modulational instability of purely stochastic origin appears. A competition between instability bands of periodic and stochastic origin is also discussed. We find an instability gain comparable to the conventional values found in a homogeneous anomalous dispersion fiber.
\end{abstract}


\maketitle

Modulational instability (MI) is a pervasive phenomenon in the physics of nonlinear dispersive waves.  It manifests itself as the  destabilization of a uniform wavepacket by the exponential growth of small harmonic perturbations around the carrier frequency of the wavepacket \cite{Zakharov2009}.  Its study originated in hydrodynamics \cite{Benjamin1967a,Zakharov1968}, but analogous phenomena were also discovered in electromagnetic waves \cite{Bespalov1966} and optical fibers \cite{Tai1986}. 
The main ingredients to observe MI are  focusing cubic nonlinearity (like the Kerr effect in silica) and  anomalous (negative) group-velocity dispersion (GVD). 

Notwithstanding, MI can be found also in normal (positive) GVD, if higher-order dispersion  \cite{Cavalcanti1991} or birefringence \cite{Biancalana2004e} are considered. 
Moreover, in single-mode fibers, the periodic variation of GVD along the fiber length can also give rise to MI in the normal GVD regime. This effect is similar to the destabilization
of a parametrically excited harmonic oscillator and is denoted as parametric MI \cite{Smith1996,Abdullaev1996,Droques2012,Armaroli2012,Mussot2018}.

Optical fibers featuring random GVD variations were also extensively studied. In the late 90s the exactly solvable white noise process was considered \cite{Abdullaev1996,Abdullaev1999,Garnier2000,Chertkov2001}. More recently some of the present authors focused on different processes such as localized GVD kicks \cite{Dujardin2021} and coloured processes of low-pass and band-pass type \cite{Armaroli2022}. 

So far, both periodic and random fluctuations {have been mostly} assumed to occur around
an average GVD  different from zero (and more often normal, to avoid competition with conventional MI, which exhibits much higher MI gain). The fluctuations can be large, though. 

On the contrary, systems with zero average GVD have attracted a lot of attention for the {suppression of the dispersion-induced pulse broadening} and the optimization of nonlinear pulse {transmission} \cite{Agrawal2012,Turitsyn2012}. {This approach is commonly denoted} as dispersion-management (DM): segments of positive and negative GVD alternate along the fiber. The study of MI in periodic DM fiber links \cite{Bronski1996c} shows a  behavior different from the parametric {MI}: a threshold is found for the segment lengths below which no MI appears. The amplitude of GVD variations influences the MI spectral range, but has no effect on this threshold.

Here we consider random fluctuations of the DM segment lengths. While pulse propagation in a similar system was analyzed in Ref.~\cite{Malomed2001}, we focus here on MI, by applying the technique developed in Ref.~\cite{Garnier2000,Dujardin2021}. After deriving some analytical relations for uniformly distributed fluctuations around the periodic arrangement, we compare them to numerical solutions. We find MI bands of purely stochastic origin and characterize the transition from  periodic to stochastic DM.

\begin{figure}[ht!]
\centering
\includegraphics[width=0.4\textwidth]{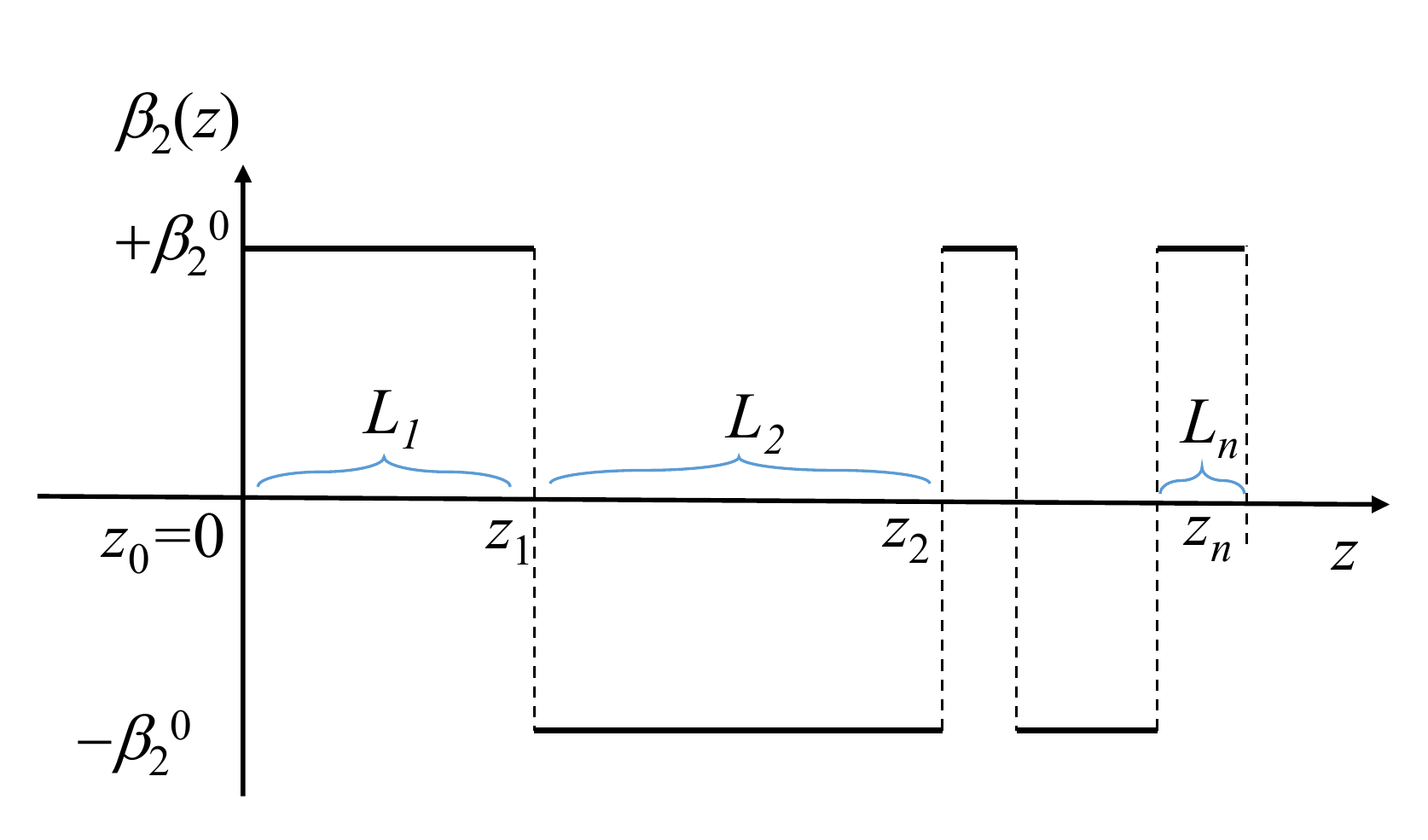}
\caption{ Schematic representation of the GVD profile in a typical fiber realisation. }
\label{fig:dispersion}
\end{figure}
{We consider the propagation of optical pulses ruled by} the nonlinear Schr{\"o}dinger equation (NLSE)\cite{Agrawal2012},
\begin{equation}
   i\partial_z U - \frac{1}{2}\beta_2(z)\partial_{tt}U  + \gamma |U|^2U = 0,
\label{eq:NLS1}
\end{equation}
where $U(t,z)$ is the complex envelope of the optical field, ($t$,$z$) are the physical time and propagation distance in a frame moving at the group velocity of the fiber mode, $\gamma$ the (constant) nonlinear coefficient, and $\beta_2(z) = \pm\beta_2^0$ ($\beta_2^0>0$) is the  GVD, which takes only two values. As schematically illustrated in Fig.~\ref{fig:dispersion}, the sign changes occur at $z_1, z_2, \ldots, z_N$, where $z_{n} = z_{n-1} +  L_n$, $n=1,2,\ldots, N$ and $z_0=0$.  The lengths $L_n$ are independent, identically distributed random variables with uniform probability distribution function in $[\bar{L}(1-\varepsilon),\bar{L}(1+\varepsilon)]$, where $\bar{L}$ is the average  length of one fiber segment (half of the DM period) and $\varepsilon$ the amplitude of the fluctuation. 

Equation (\ref{eq:NLS1}) has a continuous wave ($t$-independent) solution $U_0(z) = \sqrt{P}\exp(i \gamma P z)$. 
In order to study its stability, we insert in  \eqref{eq:NLS1} the \textit{Ansatz}   ${U}(z,t) = \left[\sqrt{P} + \check x_1(z,t) + i \check x_2(z,t)\right]\exp (i \gamma P z)$, where $\check x_{1,2}$ are assumed to be small, linearize and Fourier-transform the resulting equation with respect to $t$ ($\omega$ is used as the associated angular frequency detuning from the carrier $U_0$).  We obtain  
\begin{equation}
   \frac{\ud x}{\ud z}
	 	 = 
	 	\begin{bmatrix}
	 		0 & -g(z) \\h(z) & 0
	 	\end{bmatrix}
	 	x,
\label{eq:MIeqs1}
\end{equation}
with $x  \equiv ( x_1,x_2)^\mathrm{T}$ (functions of $ \omega$ and $z$), $g(z) = \beta_2(z) \frac{\omega^2}{2}  $ and $h(z) = g(z) +  2\gamma P$.
\eqref{eq:MIeqs1} is a system of stochastic differential equations (SDEs) for each value $\omega$.

Equation (\ref{eq:MIeqs1}) is solved in the interval $(z_0,z_N)$ as a product of random matrices, depending on the random variables $L_n$ as
\begin{align}\label{eq:product}
 x(z_N) &=  T_-(L_N)T_+(L_{N-1})\ldots T_-(L_2)T_+(L_1) x(z_0),\\
    T_{\pm}(L_n) &=
    \begin{bmatrix}
      \cos( k_{\pm} L_{n})
& -\mu_{\pm} \sin( k_{\pm} L_{n}) \\
   \mu_{\pm}^{-1}\sin( k_{\pm} L_{n})
& \cos(  k_{\pm} L_{n})
    \end{bmatrix},\label{eq:Tmatrix}
\end{align}
with $k_\pm^2 =\pm \frac{\beta_2^0\omega^2}{2}\left(\pm \frac{\beta_2^0\omega^2}{2} + 2 \gamma P\right)$, {$\mu_\pm = \pm \frac{\beta_2^0\omega^2}{2 k_\pm}$}. The sign $\pm$ is chosen according to the sign of GVD in the corresponding segment. {The wave-number} $k_+$ is always real and positive, {whereas} $k_-$ is purely imaginary in the conventional MI band $0\le \omega\le \sqrt{\frac{4\gamma P}{\beta_2^0}}$. 

\begin{figure}[ht!]
\centering
\includegraphics[width=0.45\textwidth]{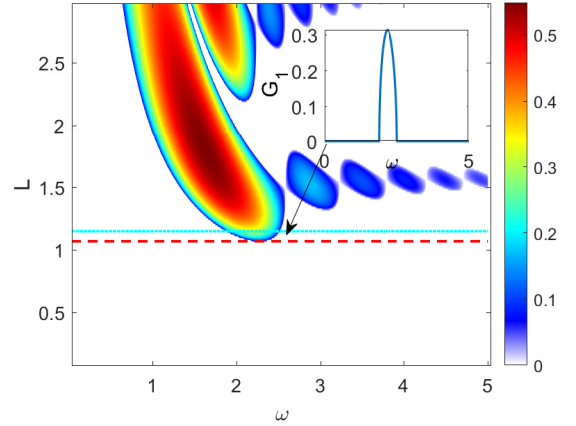}
\caption{ MI gain for a periodic DM fiber as a function of detuning $\omega$ and period $L$. The red dashed horizontal line identifies $\bar L=1.07$ (MI threshold), while the cyan dotted line denotes the reference value $\bar L=1.15$, used in the inset and below in the random length fluctuation examples.}
\label{fig:periodicmap}
\end{figure}

If the DM link is periodic, \textit{i.e.}, $\varepsilon=0$,  we can apply Floquet theory \cite{Bronski1996c}. The unit cell of DM to be periodically replicated is represented by one positive and one negative GVD trait of length $\bar L$. The MI gain is defined as {$G_1(\omega) \equiv \frac{1}{2 \bar L} \ln \max\{|\tilde\lambda|,1\}$}, where $\tilde\lambda$ is the eigenvalue of the monodromy matrix associated to \eqref{eq:MIeqs1} of largest modulus. This corresponds to $T_{2\bar L} \equiv T_-(\bar L)T_+(\bar L)$.
In \cite{Bronski1996c}, it was observed that a critical value $\bar L \approx 1.07$ exists, below which $G_1 = 0$ identically. The MI gain is represented as a false-color map in Fig.~\ref{fig:periodicmap} for $\omega\ge 0$ (for $\omega\le 0$ we obtain its mirror image). The MI gain exhibits several lobes, in general. 

For random $L_n$, in Ref.~\cite{Garnier2000,Dujardin2021} it was shown {that} we have to resort to the Lyapunov exponent of the random linear map: the sample gain 
$G_\mathrm{S}(\omega) \equiv \lim_{z_N\to \infty }\frac{1}{z_N} \ln 	\lVert T_+(L_N)T_-(L_{n-1})\ldots T_+(L_2)T_-(L_1) x(0,\omega) \rVert^2 $ converges for almost all realizations of the fiber and is a deterministic quantity.

{By taking the average of \eqref{eq:product} and letting $N\rightarrow\infty$,} we can estimate the MI gain as $G_1(\omega) \equiv \frac{1}{2 \bar L} \ln \max\left\{|\lambda|,1\right\}$, where  $\lambda$ is the largest modulus eigenvalue of $\overline T \equiv \langle T_+\rangle \langle T_-\rangle$ and the angle brackets denote the expectation operation over the random lengths $L_n$. {We can split the averages because $L_n$} are all mutually independent. Elementary integration gives
\begin{equation}
\begin{aligned}
    \langle \cos (m k_\pm n) \rangle &= \cos (m k_\pm \bar L)\frac{\sin m k_\pm \varepsilon}{m k_\pm \varepsilon}\\
    \langle \sin (m k_\pm n) \rangle &= \sin (m k_\pm \bar L)\frac{\sin m k_\pm \varepsilon}{m k_\pm \varepsilon},
   \end{aligned}
    \label{eq:avetrig}
\end{equation}
thus
\begin{equation}
\langle T_\pm \rangle =   
 \frac{\sin k_\pm \varepsilon}{k_\pm \varepsilon}
    \begin{bmatrix}
       \cos( k_{\pm} \bar L)
& -\mu_{\pm} \sin( k_{\pm} \bar L) \\
      \mu_{\pm}^{-1}\sin( k_{\pm} \bar L)
& \cos(  k_{\pm} \bar L)
    \end{bmatrix}
\end{equation}
and $\overline T$ can be easily obtained as
%
\begin{equation}
  \overline T = \frac{\sin(k_-\varepsilon)}{k_-\varepsilon}\frac{\sin(k_+\varepsilon)}{  k_+\varepsilon} T_{2\bar L}.
  \label{eq:Tav_unif}
\end{equation}
As common in random dynamical systems \cite{VanKampenBook}, we may have $G_1=0$ for all $\omega$, implying that $x_{1,2}$ decay on average.  Particularly, it is apparent that for $\bar L<1.07$, where both eigenvalues of $T_{2\bar L}$ are on the unit circle, \eqref{eq:Tav_unif} implies that $G_1(\omega)=0$ identically, because $\overline T$ differs from $T_{2\bar L}$ only by a  factor no larger than one. 

Nevertheless, a different kind of instabilitity may occur, to understand which the study of second moments is required. We let $X_1 = x_1^2$, $X_2 = x_2^2$, and $X_3 = x_1x_2$ and derive from \eqref{eq:MIeqs1}
\begin{equation}
	 	 	\frac{\ud}{\ud z}
	 	 {X} = 
	 	\begin{bmatrix}
	 		0 & 0 & -2 g(z) \\
	 		0 & 0 & 2 h(z) \\
	 		h(z) & -g(z) & 0
	 	\end{bmatrix}
	 	 {X},
	 	\label{eq:MIeqs2}
\end{equation}
with $X\equiv (X_1,X_2,X_3)^\mathrm{T}$.
\eqref{eq:MIeqs2} can be {again} solved in terms of transfer matrices $X(z_n) =  M_{\pm}(L_n) X(z_{n-1})$, with
\begin{equation}
    M_{\pm}(L_n)=
    \begin{bmatrix}
        \cos^2  k_{\pm} L_{n} &  \mu_\pm^2  \sin^2 k_{\pm} L_{n} & -\mu_\pm\sin 2 k_{\pm} L_{n} \\
        \mu_\pm^{-2}  \sin^2 k_{\pm} L_{n} &         \cos^2  k_{\pm} L_{n} & \mu_\pm^{-1}\sin 2 k_{\pm} L_{n} \\
        \frac{\mu_\pm^{-1}}{2}\sin 2 k_{\pm} L_{n} & -\frac{\mu_\pm}{2}\sin 2 k_{\pm} L_{n} & \cos 2 k_{\pm} L_{n} 
    \end{bmatrix}.
  \label{eq:Mmatrix}    
\end{equation}
The generic DM unit cell is associated to $M_-(L_{n})M_+(L_{n-1})$. After Ref.~\cite{Garnier2000,Dujardin2021}, we define {$G_2(\omega) \equiv \frac{1}{4 \bar L} \ln \max\{|\kappa|,1\}$}, where here $\kappa$ is the largest modulus eigenvalue of $\overline M \equiv \langle M_+\rangle \langle M_-\rangle$. In the periodic limit, the monodromy matrix associated to \eqref{eq:MIeqs2} is $M_{2\bar L} \equiv M_-(\bar L)M_+(\bar L)$ and $G_2=G_1$, for every $\omega$. 

{In the random case, we can easily find the expression of $\overline M$ analytically by using  \eqref{eq:avetrig}, but the expression is rather lengthy and it is not reported here}.
%
%
    
In contrast to $\overline T$, $\overline M$ is not, in general, trivially proportional to $M_{2\bar L}$. This simple algebraic consideration implies that we may have {$G_2> 0$} even when $G_1 = 0$. Therefore new MI sidebands of purely stochastic origin exist. The eigenvalues  $ \kappa$ of $\overline M$ may be found analytically, too. Their expression is very involved, though; we thus rely on a numerical routine.

In order to assess the accuracy of our estimates, we also solve \eqref{eq:MIeqs1} numerically, by taking a fixed number, $N=20$, of fiber segments.  We take $(x_1(0),x_2(0))^T=(1,0)$ [equivalently, $X(0)^\mathrm{T} = (1,0,0)$] and multiply by the transfer matrix \eqref{eq:Tmatrix} $N$ times alternating the GVD sign according to \eqref{eq:product}. We compute $P_\mathrm{out}= x_1^2(z_N) + x_2^2(z_N)$ (obviously, $P_\mathrm{in}=x_1^2(0) + x_2^2(0)=1$). We repeat this calculation taking  $N_\mathrm{iter}$ different fiber realizations.
The mean gain is defined as either 
\cite{Farahmand2004,Dujardin2021,Armaroli2022}
\begin{align}
      \overline G_1(\omega;N) & \equiv  \frac{1}{ N \bar L  }\ln \left(\left|\langle x_1(z_N)\rangle\right| +\left|\langle x_2(z_N)\rangle\right|\right)\label{eq:meangain1}\\
      \overline G_2(\omega;N) & \equiv \frac{1}{2 N \bar L  }\ln \left\langle\frac{P_\mathrm{out}}{P_\mathrm{in}}\right\rangle,\label{eq:meangain2}
\end{align}
which are compared to either $G_1$ or $G_2$, respectively.
\begin{figure}[ht!]
\centering
\includegraphics[width=0.45\textwidth]{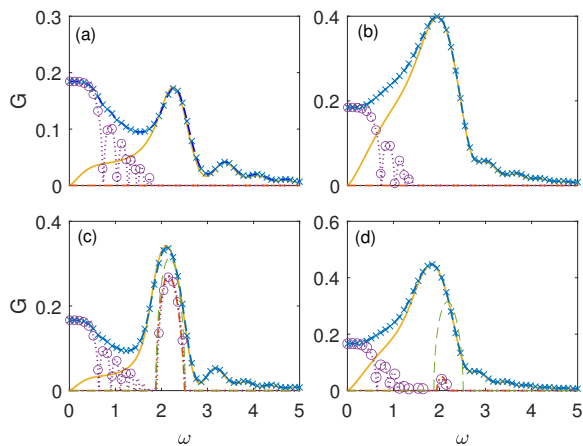}
\caption{MI gain curves for different values of $\bar L$ and $\varepsilon$, for $\omega\ge 0$. (a) $\bar L = 1$, $\varepsilon=0.2$; (b) $\bar L = 1$, $\varepsilon=0.5$; (c) $\bar L = 1.15$, $\varepsilon=0.2$;  (d) $\bar L = 1.15$, $\varepsilon=0.5$.  Blue crosses (resp.~purple circles) represent  $\overline G_2$ (resp.~$\overline G_1$) from numerical data,  the solid yellow (resp.~dash-dotted red) lines the theoretical estimates $G_2$ (resp.~$G_1$), the dashed blue (resp.~dotted purple) lines the semi-analytical estimates $\tilde G_2$ (resp.~$\tilde G_1$). In (c)-(d), we include as thin green dashed line the MI gain in the periodic case $\varepsilon=0$. }
\label{fig:MIgain}
\end{figure}

For definiteness, we take $\gamma = P = \beta_2^0 = 1$, which amounts to introduce the normalized distance $z/z_\mathrm{nl} \rightarrow z$, time $t/t_0 \rightarrow t$, and field $U/\sqrt{P} \rightarrow U$ , where $z_\mathrm{nl}=(\gamma P)^{-1}$ is the so-called nonlinear length and $t_0=\sqrt{\beta_2^0 z_\mathrm{nl}}$ is a characteristic time. The conventional MI in anomalous GVD thus reaches the maximum value of {$G_{1,\mathrm{max}} =G_{2,\mathrm{max}}=1 $} at $\omega=\sqrt{2}$; the MI sidelobes are found in $0\le|\omega|\le 2$.
We show in Fig.~\ref{fig:MIgain} four illustrative examples of MI sidelobes. 
We notice that, in general, $\overline G_2$ converges to $G_2$ for $N_\mathrm{iter}=1\times 10^6$, whereas  in general a much larger $N_\mathrm{iter}=1\times 10^7-1\times 10^8$ is required to achieve a stable and reliable estimate of $\overline G_1$.



We notice that $G_1=G_2=0$ is expected at $\omega=0$. $G_1$ exhibits a single lobe (dash-dotted red lines), while $G_2$ grows monotonically to reach a maximum value at around $\omega \approx 2$, then decays in an oscillatory way (solid yellow lines). This is not the case in  numerical results (blue crosses and purple circles), which are affected by the limited size of the numerical domain and present a finite $\overline G_{1,2}$ at $\omega=0$. This can be quantified by deriving  alternative semi-analytical estimates of gain. We notice that  $T_-(L_N) \ldots  T_+(L_1) \approx \overline T ^{N/2}$ (resp.~$M_-(L_N) \ldots M_+(L_1) \approx \overline M ^{N/2}$) and define
\begin{align}
\label{eq:G1app}
\tilde G_1 &\equiv \frac{1}{ N \bar L} \ln\left[\left(\overline T^\frac{N}{2}\right)_{11} + \left(\overline T^\frac{N}{2}\right)_{21}\right],
\\
\label{eq:G2app}
    \tilde G_2 &\equiv \frac{1}{2 N \bar L} \ln \left[ \left( \overline M^\frac{N}{2}\right)_{11} + \left(\overline M^\frac{N}{2}\right)_{21}\right],
\end{align} 
where matrix power are computed numerically and the subscripts refer to the corresponding numerically computed matrix elements. We notice in Fig.~\ref{fig:MIgain} that in every case this estimate performs very well not only for $\omega \approx 0$, but in the whole domain. 

In Fig.~\ref{fig:MIgain}(a)-(b), $\bar L = 1<1.07$ and $\varepsilon$ increases from $0.2$ to $0.5$. The MI is of purely stochastic origin. The local maximum value achieved by $\overline G_2$ at $ \omega=\omega_{2,\mathrm{max}}>0$, say $G_{2,\mathrm{max}}$, increases with $\varepsilon$ and becomes of the same order of magnitude as the conventional MI {(\textit{i.e.}, for constant anomalous GVD)}. The width of the sidelobes increases significantly as well. 

In Fig.~\ref{fig:MIgain}(c)-(d), $\bar L = 1.15>1.07$. There is thus a competition between the periodic and the stochastic effects. We include also the periodic-DM  MI sidelobe for comparison (thin green dashed line). 

For $\varepsilon=0.2$ [Fig.~\ref{fig:MIgain}(c)] the random fluctuations yield a broadened MI sidelobe. In contrast to {the case of constant anomalous GVD perturbed by white-noise}, where the broadening is accompanied by a reduction of $G_{2,\mathrm{max}}$ \cite{Abdullaev1996}, here this value is slightly enhanced. We notice also that $\overline G_1$ is always less than its periodic counterpart, consistently with \eqref{eq:Tav_unif}. Comparison of Fig.~\ref{fig:MIgain}(a) and Fig.~\ref{fig:MIgain}(c) shows that the main sidelobe appearing in the former is located in the same region of the periodic sidelobe. The random fluctuations facilitate the emergence of the MI sidelobes in a range that coincides with the periodic DM.

For a larger fluctuation ($\varepsilon=0.5$), Fig.~\ref{fig:MIgain}(d), the residual effect of periodicity is completely erased and we obtain a single wide lobe similar to the corresponding below-threshold example of Fig.~\ref{fig:MIgain}(b), but with a larger $G_{2,\mathrm{max}}$. 

\begin{figure}[ht!]
\centering
\includegraphics[width=0.45\textwidth]{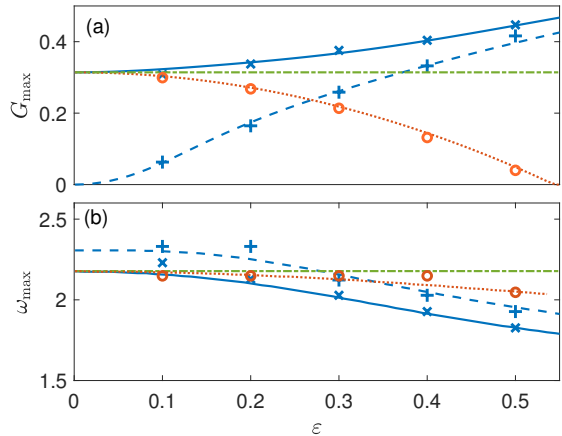}
\caption{(a) Maximum gain values and (b) their corresponding $\omega$ as a function of $\varepsilon$.  The cross (resp.~plus) markers correspond to the maxima $ G_{2,\mathrm{max}}$ for $\bar L = 1$ (resp.~ $\bar L = 1.15$). The dashed (resp. solid) blue lines correspond to the maxima of $G_2$ for $\bar L = 1$ (resp.~$\bar L = 1.15$). The green dash-dotted lines report for reference the constant values found in the periodic limit, i.e., the maxima of $G_1=G_2$ for $\varepsilon=0$. Finally the red circles show the values of $G_{1,\mathrm{max}}$ and $\omega_{1,\mathrm{max}}$ for $\bar L = 1.15$, for which red dotted lines illustrate  the corresponding maxima of $G_1$. In panel (b) the line stops at $\varepsilon\approx 0.54$ (cut-off for $G_1$).}
\label{fig:maxMIgain}
\end{figure}

In order to summarize our findings, we show in Fig.~\ref{fig:maxMIgain}(a)-(b), respectively, $G_{j,\mathrm{max}}$ and the corresponding $\omega$ value, $\omega_{j,\mathrm{max}}$, with $j=1,2$, as a  function of $\varepsilon$. Obviously, $G_{1,\mathrm{max}}$ is the local maximum value of  $\overline G_1$ at $\omega=\omega_{1,\mathrm{max}}>0$. We compare them to their corresponding theoretical estimates, \textit{i.e.}, the maxima of $G_1$ (resp.~$G_2 $). Solid blue lines, crosses, red lines and circles correspond to $\bar L=1.15$, while dashed lines and pluses correspond to $\bar L = 1$. In Fig.~\ref{fig:maxMIgain}(a) we notice that $G_{2,\mathrm{max}}$ increases monotonically with $\varepsilon$. As expected, for $\varepsilon\to 0$ the gain vanishes for $\bar L =1$ and converges to the periodic $G_{1,\mathrm{max}}=0.31$ (for $\varepsilon=0$, dash-dotted green line) for $\bar L = 1.15$. For large $\varepsilon$ they converge to two very similar values, which is around 30$\%$ smaller than the conventional MI value.
 The dotted red line in Fig.~\ref{fig:maxMIgain}(a) shows the maxima of $G_1$ obtained from \eqref{eq:Tav_unif}. For $\varepsilon > 0.54$, $G_1=0$ for every $\omega$. For this value of the fluctuation amplitudes, randomness completely overrule the effects of periodicity. This is corroborated by the values ${G}_{1,\mathrm{max}}$ (red circles), which match very well with the theoretical estimates (provided that a large enough $N_\mathrm{iter}$ is chosen). 
 In Fig.~\ref{fig:maxMIgain}(b) we observe that for both $\bar L$, $\omega_{2,\mathrm{max}}$ decreases with $\varepsilon$ and converges to a value larger than the conventional MI value. It is always below its periodic counterpart for $\bar L = 1.15$, apart from numerical fluctuations. We also notice that ${ \omega}_{2,\mathrm{max}}<{ \omega}_{1,\mathrm{max}}<2.18$, \textit{i.e.}, slightly below the $\varepsilon=0$ value. For $\bar L = 1$, $\omega_{2,\mathrm{max}}$ is  above its periodic counterpart and crosses it for  $\varepsilon \approx 0.3$. The theoretical estimates work very well for every considered value of $\varepsilon$.

To conclude, we investigated the effect of uniformly distributed random fluctuations of the \emph{length} of DM fiber links. We considered the MI problem and {developed} an analytical technique to estimate the instability gain. 
The MI gain attains values comparable with the conventional ones in a homogeneous anomalous GVD fiber and up to 50$\%$ larger those found for the periodic arrangement. Comparison to the direct numerical solution in a Monte Carlo fashion confirm the soundness of the method.

This {may be of} of interest for tailoring and control of MI sidebands for telecommunications and parametric sources.

\section*{Acknowledgments}
The present research was supported by IRCICA (USR 3380 CNRS), Agence Nationale de la Recherche (Programme Investissements d’Avenir, I-SITE VERIFICO, Labex CEMPI); Ministry of Higher Education and Research; Hauts de France Council; European Regional Development Fund (Photonics for Society P4S, Wavetech), CNRS (IRP FELANI).

\bibliography{Randomfibers}

\end{document}